\documentclass[12pt]{article}
\usepackage {amsmath}
\usepackage[dvips]{graphicx}
\usepackage{feynmp}
\usepackage{amssymb}
\usepackage{cite}
\newcommand{\be}{\begin{equation}}
\newcommand{\ee}{\end{equation}}
\newcommand{\bear}{\begin{eqnarray}}
\newcommand{\eear}{\end{eqnarray}}

\newcommand{\lapproxeq}{\lower .7ex\hbox{$\;\stackrel{\textstyle  
<}{\sim}\;$}} 
\newcommand{\gapproxeq}{\lower .7ex\hbox{$\;\stackrel{\textstyle  
>}{\sim}\;$}} 
\newcommand{\stackdown}[2]{\lower 1.4ex\hbox{$\;\stackrel{\textstyle{#1}}  
{\scriptstyle{#2}}\;$}}
\newcommand{\beq}{\begin{equation}} 
\newcommand{\eeq}{\end{equation}} 

\newcommand{\ba}{\begin{eqnarray}}
\newcommand{\ea}{\end{eqnarray}}

\newcommand{\bea}{\begin{eqnarray}}
\newcommand{\eea}{\end{eqnarray}}

\makeatletter 
\def\slash{\@ifnextchar[{\fmsl@sh}{\fmsl@sh[0mu]}} 
\def\fmsl@sh[#1]#2{%
  \mathchoice 
    {\@fmsl@sh\displaystyle{#1}{#2}}%
    {\@fmsl@sh\textstyle{#1}{#2}}%
    {\@fmsl@sh\scriptstyle{#1}{#2}}%
    {\@fmsl@sh\scriptscriptstyle{#1}{#2}}} 
\def\@fmsl@sh#1#2#3{\m@th\ooalign{$\hfil#1\mkern#2/\hfil$\crcr$#1#3$}} 
\makeatother 
\begin{document}
\begin{titlepage}  
\begin{flushright} 
\parbox{4.6cm}{UA-NPPS/BSM-04/13 }
\end{flushright} 
\vspace*{5mm} 
\begin{center} 
{\large{\textbf {Universe Acceleration in Brane World Models.
}}}\\
\vspace{14mm} 
{\bf C. ~\ Chiou-Lahanas}, \,{\bf G. A.~\ Diamandis} and {\bf B. C.~\ Georgalas} 
{\footnote{email: \, cchiou@phys.uoa.gr, \, gdiam@phys.uoa.gr, \, vgeorgal@phys.uoa.gr}}
\vspace*{6mm} \\
  {\it University of Athens, Physics Department,  
Nuclear and Particle Physics Section,\\  
GR--15771  Athens, Greece}

\end{center} 
\vspace*{25mm} 
\begin{abstract}
We examine the cosmology induced on a brane moving in the background of a five-dimensional black hole, solution of the string effective action. The evolution, determined by the Israel junction conditions is found to be compatible with an accelerating universe with the present day acceleration coming after a decelerating phase. The possible species of the  energy-momentum tensor, localized on the brane, for these solutions to be valid are discussed. 

\end{abstract} 
\end{titlepage} 
\newpage 
\baselineskip=18pt 
\section*{Introduction}

The idea that the visible world may be realized as a brane in a higher dimensional space, where matter is confined is not new \cite{Rubakov, Akama, Visser1, Gibbons1}.  A  natural framework for this idea is String Theory in which 
higher dimensional spacetime arises by fundamental requirements.  The recognition that not all extra dimensions are of the order of Planck scale \cite{Antoniadis} and the discovery of D-branes \cite{Polchinski}, naturally 
existing in string theory, revived this idea in  recent years. The brane world scenarios  \cite{review} have been extensively studied in order to address  fundamental questions such as the 
hierarchy problem \cite{Antoniadis1, Antoniadis2 , Randall, Randall1, Gogbe, Gogbe1}, the cosmological constant problem and issues regarding the early time cosmology.  In a large class of these models a Friedman-Robertson-Walker universe arises by the evolution of a three-dimensional brane in an AdS  or in an AdS black-hole spacetime \cite{cosmo, Binetruy, cosmo1, cosmo2, Chamblin, cosmo3, cosmo4, Flanagan, cosmo5, cosmo6, cosmo7, deAlwis, Kehagias1, Grojean, Kraus, Barcelo, Kehagias, Papantonopoulos,  Cai, Kanti, Cho, BouhmadiLopez, Neupane1, Neupane2}. 
Certainly the origin of the bulk Lagrangian and its relation to an underline fundamental theory is always under investigation. Towards this direction  higher dimensional black hole solutions have been extensively studied in various cases and especially in the context of Supergravity 
and String Theory \cite{Gibbons, GHS, Duff, Duff1, Yazad, Aliev, Buchel, Castro}.
 
In a previous work \cite{Chiou} we examined  background solutions of the string effective action in five dimensions. In particular we derived the 
unique dilaton - metric static vacuum with non-vanishing Liouville-type dilaton potential. This solution represents a Schwarzschild type black hole.  
It was also found that this solution can be extended to incorporate electric charge leading to charged five-dimensional black hole. Furthermore 
Kalb-Ramond charge may be also present. In this case the three-dimensional space has to be closed. The thermodynamic properties of these black hole backgrounds were studied and especially their masses and their temperatures were calculated.  Moreover we presented a brief discussion of the cosmology induced on a brane evolving in this background introducing  a single tension term on the brane. 
We found that the present day acceleration of the universe may have the observed value and that the accelerating phase follows a decelerating one with the transition occurring to phenomenologically acceptable value of the redshift \cite{Shapiro}. Nevertheless the consideration of just the brane tension was not adequate to match the Israel junction conditions on the brane. In this work we see that introducing more general energy momentum tensor on the brane the behavior of the acceleration  is maintained and the junction conditions are satisfied.
\section*{Five-Dimensional String Effective Action Background}
In this section we briefly review the derivation of the static five-dimensional background found in \cite{Chiou} which will be used for the cosmological study.
The bosonic part of the low energy limit of the string  effective action is
\be
S\,=\, \frac{1}{2\kappa _0^2} \int d^Dx \sqrt{-G} e^{-2 \tilde{\phi} } \left \{ R\,+\,4(\nabla \tilde{\phi} )^2\,-\,
\frac{1}{12}H_{\mu \nu \rho }H^{\mu \nu \rho } \,-\, \frac{1}{4} F_{\mu \nu }F^{\mu \nu } \,-\,\Lambda  \right \}
\label{sfaction}
\ee
where $\Lambda$ is a constant related to the central charge. In particular $\Lambda = \frac{2(D-26)}{3\alpha '}$ for the bosonic string, and $\Lambda = \frac{(\frac{3D}{2}-15)}{3\alpha '}$ for the superstring, 
  $\tilde{\phi} $ is the dilaton field and
 $H_{\mu \nu \rho }\,=\, \nabla _{\mu } B_{\nu \rho }+ \nabla _{\rho  } B_{\mu \nu  } + \nabla _{\nu } B_{\rho \mu  }$ is the field strength of the antisymmetric tensor field $B_{\mu \nu }$. We 
also assume the existence of $U(1)$ gauge field  $A_{\mu}$ with $F_{\mu \nu }=\partial _\mu A_\nu -\partial _\nu A_\mu $ its field strength. 
Such a gauge field can be considered as a remnant of a partial compactification.\\
The curvature term can be cast to the Einstein-Hilbert form by the Weyl rescaling of the metric
$$
g_{\mu \nu }\,=\, e^{-\frac{4\phi }{D-2}} G_{\mu \nu }\, \quad \phi \, = \, \tilde{\phi} - \phi_0.
$$ 
Doing this  the dilaton appears as a matter field with a Liouville-type potential and of particular couplings with the other fields
\be
S\,=\, \frac{1}{2\kappa ^2} \int d^Dx \sqrt{-g}  \left \{ R\,-\,\frac{4}{D-2}(\nabla \phi )^2\,-\,
\frac{1}{12}e^{-\frac{8\phi }{D-2}}H^2 \,-\, \frac{1}{4}e^{\frac{-4\phi }{D-2}} F^2 \,-\, V(\phi ) 
 \right \}
\label{efaction}
\ee
where $V(\phi )\,=\,\Lambda  e^{\frac{4\phi }{D-2}}$. Note that $\kappa ^2 =  e^{2 \phi_0} \kappa_0^2 = (8 \pi G_D)$ is related to the D-dimensional Newton's constant.

For a five-dimensional space-time the field equations read
\bea
R_{\mu \nu }-\frac{1}{2}g_{\mu \nu }R &=& \sigma _1 \left( \nabla _{\mu } \phi \nabla _{\nu }
\phi - \frac{1}{2}g_{\mu \nu } (\nabla \phi)^2 \right) - \frac{1}{4} e^{\sigma _2 \phi } 
\left ( H_{\mu \rho \sigma }H_{\nu }^{\rho \sigma } - \frac{1}{6}g_{\mu \nu}H^2 \right ) \nonumber \\
 &+& \frac{1}{2}e^{\sigma _4\phi } \left ( F_{\mu \rho } F_{\nu }^{\rho } - \frac{1}{4} g_{\mu \nu } F^2 \right ) - \frac{\Lambda }{2} e^{\sigma _1\phi }g_{\mu \nu } \nonumber
\\
2 \Box \phi - \Lambda e^{\sigma _1 \phi }&+& \frac{1}{6}e^{\sigma _2 \phi }H^2 + \frac{1}{4}e^{\sigma _4 \phi }F^2 = 0  \nonumber \\
\nabla _ \rho  \left \{ e^{ \sigma _2 \phi } H ^\rho _{\mu \nu } \right \} &=& 0 \nonumber \\
\nabla _ \mu  \left \{ e^ {\sigma _4 \phi } F^{\mu \nu } \right \} &=& 0 
\eea
where $\sigma _1=-\sigma _4=4/3\, \quad \sigma _2= -2 \sigma_1 = -8/3$. \\
Setting
\be
H^{\mu \nu \rho } = \frac{1}{2} e^{- \sigma _2 \phi } E^{\mu \nu \rho \sigma \tau }
\mathcal{B}_{\sigma \tau }\,,
\label{dual}
\ee
where $E^{\mu \nu \rho \sigma \tau }$ is the totally antisymmetric tensor in five dimensions and
$ \mathcal{B}_ {\sigma \tau } = \partial _{\sigma } \mathcal{K}_\tau - \partial _{\tau  } \mathcal{K}_ \sigma $ is the field strength of a vector field $\mathcal{K}_{\mu }$, the equations become
\bea
R_{\mu \nu }-\frac{1}{2}g_{\mu \nu }R &=& \sigma _1 \left( \nabla _{\mu } \phi \nabla _{\nu } \phi - 
\frac{1}{2}g_{\mu \nu } (\nabla \phi)^2 \right) + \frac{1}{2} e^{-\sigma _2 \phi }
 \left ( \mathcal{B}_{\mu \rho  } \mathcal{B}_{\nu }^{\rho } - \frac{1}{4}g_{\mu \nu } \mathcal{B}^2 \right ) \nonumber \\
 &+& \frac{1}{2}e^{\sigma _4\phi } \left ( F_{\mu \rho } F_{\nu }^{\rho } - \frac{1}{4} g_{\mu \nu } F^2 \right ) - 
\frac{\Lambda }{2} e^{\sigma _1\phi }g_{\mu \nu } \nonumber 
\\
2 \Box \phi - \Lambda e^{\sigma _1 \phi }&-& \frac{1}{2}e^{-\sigma _2 \phi }\mathcal{B}^2 + 
\frac{1}{4}e^{\sigma _4 \phi }F^2 = 0 \nonumber  \\
\nabla _ {\mu  } \left \{ e^ {-\sigma _2 \phi } \mathcal{B} ^{\mu \nu } \right \} &=& 0 \nonumber  \\
\nabla _ \mu  \left \{ e^ {\sigma _4 \phi } F^{\mu \nu } \right \} &=& 0 \,.
\label{dualeq}
\eea
Note that the substitution (\ref{dual}) interchanges the equation of motion with the Bianchi identity for 
the Kalb-Ramond field and the equations 
(\ref{dualeq}) are those derived from the action
\be
S\,=\, \frac{1}{2\kappa ^2} \int d^5x \sqrt{-g}  \left \{ R\,-\,\frac{4}{3}(\nabla \phi )^2\,-\,
\frac{1}{4}e^{-\sigma _2 \phi } \mathcal{B}^2 \,-\, 
\frac{1}{4}e^{\sigma _4} F^2 \,-\, V(\phi )  \right \}\,.
\label{dualaction}
\ee

The general spherically  symmetric metric-dilaton string background, under the assumption for a static dilaton, is uniquely determined to be a Schwarzschild  type black hole of the form \cite{Chiou, Kiritsis}:

\be
ds^2 \,=\, -y^2U(y)dt^2 + \frac{dy^2}{U(y)} + y^2d\Omega _3^2
\label{line}
\ee
where 
\[
U(y) = \lambda - \frac{M}{y^3} .
\]
The three-dimensional space is  flat
\be
d \Omega ^2_3  
\,=\,
dr ^2+r^2  (d \theta ^2 + sin^2 \theta d \varphi ^2) \,.
\ee
The positive parameter $\lambda $ is related to the central charge deficit for the fermionic or the bosonic string, and $M$ gives the mass of the black hole. The event horizon is located at 
\[
y_H \,=\, \left( \frac{M}{\lambda } \right)^{(1/3)}\, 
\]
and the physical singularity at $y \,=\, 0$. The dilaton field is given by
\[
\phi \,=\, m - \frac{3}{2}\,ln\,y
\]
and in this form  the dilaton drives the string coupling $e^{\phi }$ to a divergent value at the physical singularity.
    
\section*{Induced Cosmology on a Moving Brane}



In order to study the induced cosmology \cite{cosmo, Binetruy, cosmo1, cosmo2, Chamblin, cosmo3, cosmo4, Flanagan, cosmo5, cosmo6, cosmo7, deAlwis, Kehagias1, Grojean,  Kraus, Barcelo, Kehagias, Papantonopoulos,  Cai, Cho, BouhmadiLopez, Neupane1, Neupane2, Brax, Kanti1, Tetradis} we introduce a brane at the point $y \,=\, R$. 
The world volume spanned by the brane is determined by the vector $X^{\mu }\,=\,(t(\tau ),\, R(\tau ),\, x^i)$ where $\tau $ is the proper time. 
The five-dimensional line element is of the form
\be
ds_5^2 \,=\, G_{\mu \nu } dx^\mu  dx^\nu \,=\, -A(y) dt^2 \,+\, B(y) dy^2 \,+\, y^2 \,d\Omega _3^2\,,
\ee
where $A, \,\,B\,\,$ are given in equation \ref{line}.
The metric induced on the brane, $g_{ab}\,=\, \frac{\partial X^{\mu }}{\partial \xi  ^a} \, \frac{\partial X^\nu }{\partial \xi  ^b} \, G_{\mu \nu }$, where $\xi ^a \,=\, (\tau ,\,x^i)$ yields the line element
\be
ds_{4}^2 \,=\, -d\tau ^2 + R^2(\tau ) d\Omega _3^2\,,
\label{fourmetric}
\ee
provided that $\frac{dt}{d\tau } \,=\, \sqrt{\frac{1+B \dot{R}^2}{A}}$. The velocity vector is 
$u^{\mu } \,=\, (\frac{dt}{d\tau },\, \dot{R},\, \vec{0})$ where dot denotes the derivative with respect to the proper time $\tau $.
The unit vector normal to the brane directed inside the bulk space described by the original five-dimensional metric is uniquely determined by the conditions $u_\mu  \eta ^\mu \,=\,0$ and $\eta _\mu  \,\eta ^\mu \,=\,1 \,,$ given by
$\eta ^{\mu } \,=\, \frac{1}{\sqrt{AB}}\,(-B\,\dot{R},\, -A\, \frac{dt}{d\tau },\, \vec{0}) $ .
Assuming that the brane describing the visible spacetime is the boundary separating two regions of the five-dimensional spacetime with two different metrics the Israel junction conditions impose that
\be
K_{ab }^+ - K_{ab }^- \,=\, - 8\pi G \left[ T_{ab} - \frac{1}{3}\,T\,g_{ab}\right]
\ee
where 
\be
K_{ab} \,=\, \frac{1}{2} \frac{\partial X^{\mu }}{\partial \xi  ^a} \, \frac{\partial X^\nu }{\partial \xi  ^b} \, (\eta _{\mu ; \nu } 
\,+\, \eta _{\nu ;\mu }) \ee
  is the extrinsic curvature of the brane, $T_{ab}$ is the energy momentum tensor on the brane and $T$
its trace. Imposing a $Z_2$ symmetry we get
\be
K_{ab }\,=\, - 4\pi G \left[ T_{ab} - \frac{1}{3}\,T\,g_{ab}\right].
\ee

In our case the components of the extrinsic curvature are
\bea
K_{ij} &\,=\,& -  \frac{1}{R} \, \frac{\sqrt{1 + \dot{R}^2B}}{\sqrt{B}} \,  g_{ij}  \\
K_{\tau \tau } &\,=\,& \frac{1}{\dot{R} \sqrt{AB}} \, \frac{d}{d\tau } \,\left\{ \sqrt{A}\, \sqrt{1 + \dot{R}^2 B}\right \} \,.
\eea

Assuming that on the brane we have

\be
T_{ab} \,=\, (-\rho, \,\,p, \,\,p, \,\,p)
\ee
and imposing $Z_2$ symmetry, the junction conditions lead to the equations 
\be
\frac{1}{R} \, \frac{\sqrt{1 + \dot{R}^2B}}{\sqrt{B}} \,=\, \frac{4 \pi G}{3} \rho 
\label{junction1}
\ee
for the space components and 
\be
\frac{1}{\dot{R} \sqrt{AB}} \, \frac{d}{d\tau } \,\left\{ \sqrt{A}\, \sqrt{1 + \dot{R}^2 B}\right \} \,=\, - \frac{4 \pi G}{3} (2\rho  \,+\, 3p)
\label{junction2}
\ee
for the time component. 

The  equation (\ref{junction1}) is easily recognized  as a Friedman-like equation for the brane cosmology with the known peculiarity for the Hubble parameter being analogous to $\rho $ instead of $\sqrt{\rho }$ namely 
\be
H^2 \,=\, \frac{\dot{R}^2}{R^2} \,=\, \left(\frac{4\pi G}{3} \rho (R) \right)^2 \,-\, \frac{U(R)}{R^2}.
\label{hubble1}
\ee
where $G$ is the five-dimensional gravitational constant.  Introducing
$\rho _{cr} \,=\, 3H_0^2/(8\pi G_N)$, as is defined in standard cosmology and rescaling with the Hubble constant $H_0$  the equation above becomes
\be
\tilde{H}^2 \,=\,  \tilde{\rho }^2 \frac{G^2 H_0^2}{4 G_N^2} \,-\, \tilde{U}.
\label{crith}
\ee
where $\tilde{\rho } = \rho / \rho _{cr}$ and and $\tilde{U}\,=\, U/(R^2 H_0^2).$
From this  relation   the four dimensional gravitational constant $G_N$ and the five-dimensional one $G$ are related as
\be
G \,=\, 2 \sqrt{1+\tilde{U}(R_0)} \,G_N H_0^{-1}
\ee 
if the present day value $\rho $  equals to $\rho _{cr}$. This means that for this class of models $G$ becomes strong \cite{strong}.

 Substituting the factor $\sqrt{1 + \dot{R}^2B}$ from  equation  (\ref{junction1}) to the second (\ref{junction2}) a conservation-like equation is obtained for the energy density $\rho $, namely

\be
\dot{\rho } \,+\, 3(\rho + p) \frac{\dot{R}}{R} \,+\,  \frac{\rho }{2} \frac{d}{d\tau } ln (AB) \,=\,0.
\label{conserv}
\ee

The above equations are sufficient  to solve the cosmological model. 

Note that in the case that $AB \,=\, const$ the equation (\ref{conserv}) is just the conservation equation for ordinary matter on the brane. On the contrary in our case that $AB$ is not constant the junction condition is not satisfied if ordinary species of energy momentum tensor  obey the normal state equation that is $ p _i \,=\ w_i \rho _i$ where $i \,=\, d, r, \sigma $ stands for non-relativistic matter (dust), radiation and cosmological constant and $w_i \,=\, 0, \, 1/3, \, -1$ respectively. Indeed if we insist that 
\be
\dot{\rho} _d \,+\, 3 \frac{\dot{R}}{R}\,\rho _d \,=\, 0, \quad  \dot{\rho} _r \,+\, 4 \frac{\dot{R}}{R}\,\rho _r \,=\, 0, \quad \dot{\rho }_{\sigma } \,=\, 0
\ee
then (\ref{conserv}) gives  $ (\rho_b + \rho _r + \sigma  )\frac{1}{\sqrt{AB}} \frac{d}{d\tau } \sqrt{AB} \,=\,0$
compatible only with  static universe.  In this work we will try to make the junction conditions to be satisfied,  keeping the energy conservation on the brane. That is we do not consider bulk-brane energy exchange. In this procedure we examine two ways in achieving this goal,  while keeping the usual form of the state equations for the ordinary  matter and radiation.  Our main concern is  the acceleration parameter
 given by
\be
q \,\equiv \, \frac{R \ddot{R}}{\dot{R}^2} \,=\, \frac{R}{2 H^2} \frac{d}{dR} H^2 \,+\, 1.
\ee
We seek solutions giving the present day value of $q$ and a phenomenologically acceptable value of the redshift $z$ where the transition from the decelerating  phase occured.

{\bf {a.}} In the first case we  keep the brane tension, $\tilde{\rho }_\sigma = const$, and we introduce a sort of {\it{exotic}} matter with energy density $\tilde{\rho} _e$ satisfying a state equation $\tilde{p}_e \,=\, w_e \tilde{\rho} _e$ where the constant $ w_e$ is a free parameter.  Its role is to restore the consistency of the junction conditions. Similar consideration was adopted for other reasons in \cite{case}. Considering now
\[
\tilde{\rho}  \,=\, \tilde{\rho} _{\sigma}  \,+\, \tilde{\rho} _d \,+\, \tilde{\rho} _r \,+\, \tilde{\rho} _e
\]
the conservation equation coming from the junction condition becomes
\be
\frac{d\tilde{\rho} _e}{dR} \,+\, \frac{4+3w_e}{R}\tilde{\rho} _e \,+\, \frac{\tilde{\rho} _{\sigma}  + \tilde{\rho} _d + \tilde{\rho} _r}{R} \,=\,0.
\ee
This is easily solved and the solution as a function of the redshift is
\be
\tilde{\rho} _e \,=\, c_0 (1+z)^{4+3w_e} - \frac{\tilde{\rho }_d^0}{1+3w_e}(1+z)^{3} - \frac{\tilde{r }_d^0}{3w_e}(1+z)^{4} -  \frac{\tilde{\rho }_{\sigma }}{4+3w_e}
\label{exotic}
\ee
where the superscript denotes the present day value and the constant $c_0$ is related to $\tilde{\rho} _e^0$ in an obvious manner.
The general feature of the solutions satisfying 
 $\tilde{\rho }_0 = 1,$  $q(0) = 0.61$ and with phenomenologically acceptable $\tilde{\rho} _d$ and $\tilde{\rho }_r$ is that  both $\tilde{\rho} _e^0, \, w_e$ have to be negative and in particular  $-1 < w_e < 0$. That is this {\it{exotic}} matter is quintessence-like \cite{quint}. The new feature here is that it coexists with a constant brane tension term. It is interesting to note that the vanishing of the acceleration occurs also in an acceptable value of the redshift. In figure 1 we show the acceleration as a function of $z$ in a solution where for simplicity the position of the brane today is taken to be on the horizon, the values of $\tilde{\rho} _d^0$ and $\tilde{\rho }_r^0$ are 
 0.31 and 0.0005 respectively and we chose for $w_e $ the more phenomenologically favored value of the quintessence models -0.6 \cite{quintw}. The values of $\tilde{\rho} _\sigma $ and $\tilde{\rho} _e^0$ are determined from the conditions to be 1.97 and 
 -1.28 respectively. The acceleration turns out to vanish at z = 0.21.
\begin{figure}
\begin{center}
\vspace{0.5cm}
\includegraphics[width=10cm, height=8cm]{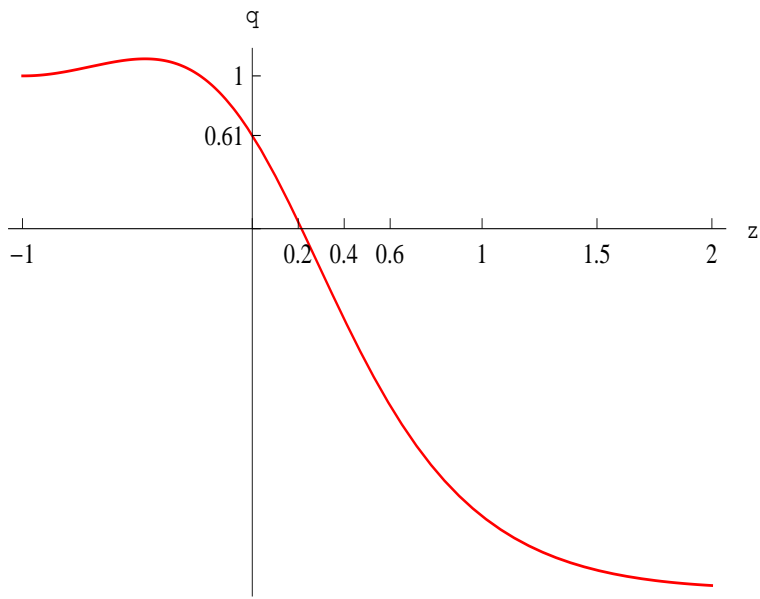}
\end{center}
\vspace{0.5cm}
\caption{ The acceleration as a function of $z$. The transition  from the decelerating to the accelerating phase occurs  at $z=0.21$.}
\label{Figure 1}  
\end{figure}

{\bf {b.}} In the second case we set $\tilde{\rho} _{\sigma} = 0$, so we are left with $\tilde{\rho} _e$  which now satisfies the equation 
\be
\frac{d\tilde{\rho} _e}{dR} \,+\,  \frac{(4
\,+\,3w_{e})}{R} \,+\,  \frac{\tilde{\rho} _d \,+\, \tilde{\rho} _r}{R} \,=\, 0
\ee
yielding
\be
\tilde{\rho} _e  \,=\,  c _0 (1+z)^{4+3w_e } \,-\, \frac{\tilde{\rho} _d^0}{3w_e + 1} (1+z)^3 \,-\, \frac{\tilde{\rho}_r^0}{3w_e} (1+z)^4 
\label{varsigma}
\ee
Solutions satisfying our requirements can  be found in this case also provided that $w_e$ is close to $-4/3$. That is we have a state equation of phantom-like matter \cite{phantom}. We see from equation (\ref{varsigma}) that if $w_e = -4/3$, $\tilde{\rho }_e$ consists of a constant term plus terms casting for the modification of the continuity equation. So if $w_e$ is close to $-4/3$,  $\tilde{\rho }_e$ has no constant term but the first term in (\ref{varsigma}) is slowly varying over finite ranges of $R$. Besides the redshifts at which the acceleration vanishes are lower than in the first one.   In figure 2 we give the $z-$ dependence of the acceleration parameter with typical values of the quantities  involved. 

Comparing the two cases some comments are in order. In the first case the presence of the brane tension makes easy the present data to be met. The second case shares in general the same features but it is more restrictive. For example the present day position of the brane is not comletely free. It has to be far from the black-hole horizon. This is not necessarily a disadvantage since the brane moves in the bulk. The real disadvantage of this case is that we cannot achieve large values of the present day dust density, in fact $\tilde{\rho }_d < 0.23$ if we insist in acceptable values of the deccelaration parameter and the redshift value of the transition. So this case seems to be disfavored by the data. Nevertheless let us point out that  in this treatment we do not consider brane-bulk energy exchange, so from our point of view, that the main concern is the present day acceleration to follow a deccelerating phase, the two cases are of similar importance. Another feature of the second case is that the early universe is radiation dominated while in the first case $ \tilde{\rho}_{e} $ may exceed $ \tilde{\rho}_{r} $ at early times if $w_e > -1/3$. In figure 3 we give the $z-$dependence of the various species of the energy density for the first case which in our treatment is more realistic. Note that  $w_e $ chosen gives a radiation dominated universe at early times.
\begin{figure}
\begin{center}
\vspace{0.5cm}
\includegraphics[width=10cm, height=8cm]{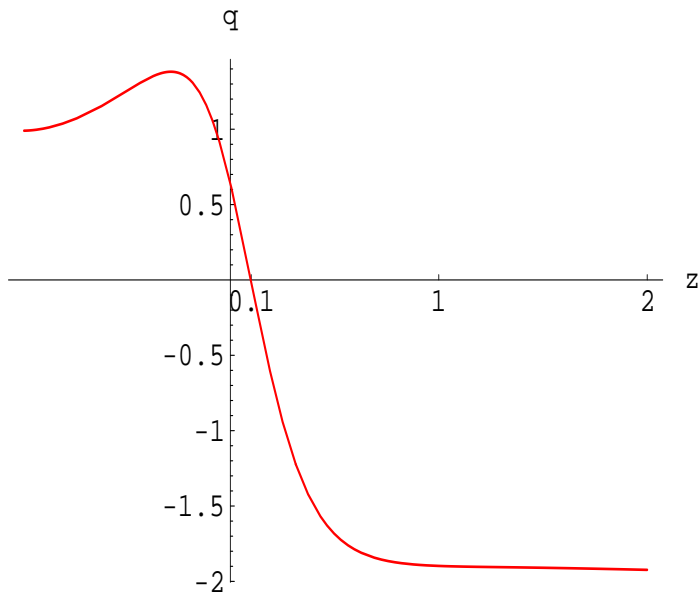}
\end{center}
\vspace{0.5cm}
\caption{ The acceleration as a function of $z$. The transition from the decelerating to the accelerating phase occurs at $z\simeq 0.1$. For this plot we have taken $w_{e } = -1.33$,  $ \tilde{\rho}_{e }(0)=0.80$ and $\tilde{\rho}_{d}(0) = 0.20 $.  The present day position of the brane is at $R(0) = 8 R_H$.}
\label{Figure 2}  
\end{figure}
\begin{figure}
\begin{center}
\vspace{0.5cm}
\includegraphics[width=10cm, height=8cm]{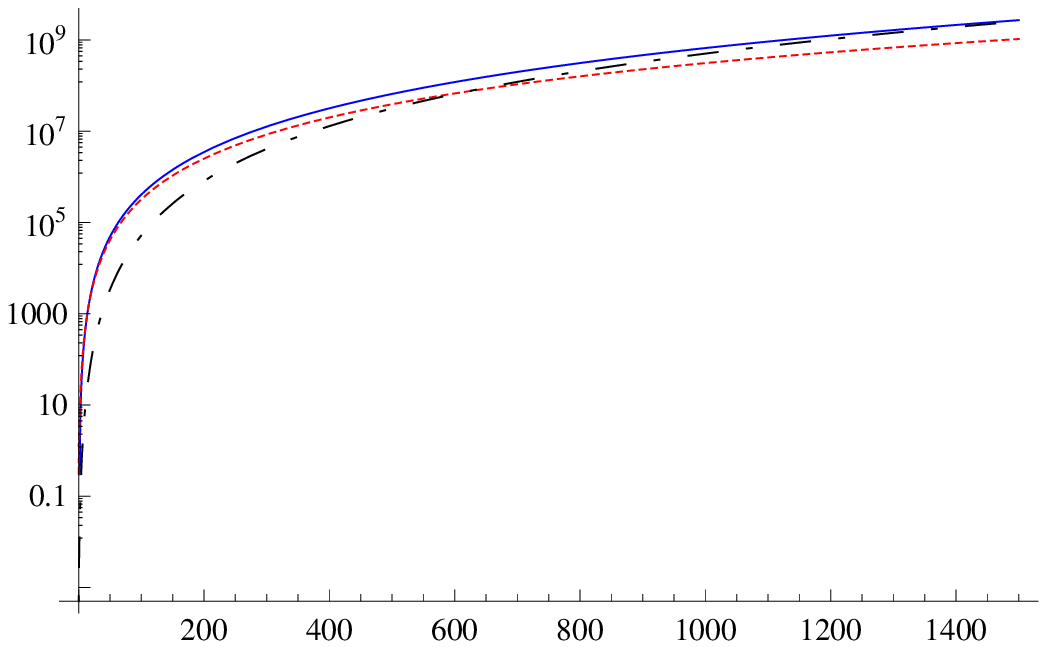}
\end{center}
\vspace{0.5cm}
\caption{ The $z-$dependence of the various species of the energy momentum tensor. The blue (solid) red (dotted) and black (dotted-dashed) lines represent $\tilde{\rho}_{e}, \, \tilde{\rho}_{d}$ and $\tilde{\rho}_{r}$ respectively. The two crossing points leading to a radiation dominated universe at early times are shown .}
\label{Figure 3}  
\end{figure}

\section*{Summary and Discussion}

  In this work we examine the cosmology induced on a brane moving in the five-dimensional Schwarzschild stringy black hole  background found in \cite{Chiou}, paying attention mainly in the acceleration parameter and insisting in deriving the today value of the acceleration to follow  a decelerating phase. Indeed we found that introducing on the brane matter with the characteristics of quintessence  or phantom even in the simple case we have considered with constant $w_e$ and no coupling of the brane matter to the dilaton field, acceptable behavior of the acceleration parameter can be achieved.
  
  This approach shares basic features with four dimensional considerations of dilaton cosmology inspired from string theory and from non-critical strings. In particular the modification to the continuity equation of brane the energy density  due to the form of the five dimensional background is  given by $\rho \frac{1}{\sqrt{AB}} \frac{d}{d\tau } \sqrt{AB}$ as is seen from eq. (\ref{conserv}). Taking into account the value of the dilaton field on the brane which is now time dependent this is just $-\frac{2}{D-2}\rho \dot{\phi }$. Note that this term for $D=4$ becomes $-\rho \dot{\phi }$ coinciding with the modification to the continuity equation due to the presence of the dilaton field in four-dimensional considerations studied in \cite {nikos1, nikos2, lahanas2, lahanas3}. In the first of  the above cases the dilaton field as is seen from the brane tends asymptotically to the linear dilaton vacuum  if it is re-expressed in terms of the five-dimensional time. Moreover since the considered black hole is asymptotically Rindler we see that we are driven to a conformal solution that is flat space with linear dilaton \cite{case, aben}. In particular from the equations (\ref{fourmetric}) and 
(\ref{hubble1}) we have that  $dy/d\tau  = y \sqrt{\rho ^2 - U/y^2}$ and $dt/d\tau = \sqrt{H^2 + U/y^2}/U$. So in the presence of brane tension where $\rho $ and $H$ tend to a constant value we have that asymptotically $dy/d\tau  \sim (const)y$ and $dt/d\tau \sim (const)$ that is $t \sim (const)lny$. This implies that $\phi (y(\tau )) \sim -Q t(\tau )$, with $Q$ a positive constant. Moreover the five-dimensional space-time  metric asymptotically is conformally  flat in the Einstein frame $ds_E^2 \longrightarrow  -y^2dt^2 + dy^2 + y^2 d\Omega _3^2   \,=\, y^2 \left(-dt^2 + dz^2 + y^2 d\Omega _3^2  \right)$ where $y dz = dy$. So in the string frame $ds^2_s = e^{4\phi/3 } ds_E^2$ the metric becomes flat. 

Finally let us note that the five-dimensional solution used in this work has as effect the necessity of incorporating types of exotic matter. No other characteristics of the black-hole solution are involved in the brane cosmological evolution. Such an involvement should arise if bulk-energy exchange occurs, a situation not considered in this work. The introduction  on the brane, dilaton-dependent energy-momentum tensor will imply additional modification to the continuity equation and one more junction condition for the dilaton field. In fact in this case we have bulk-brane energy.  This exchange occurs also in the case that the five dimensional metric is not static. The study of this cases, which should probably reveal a connection of the black-hole characteristics to the induced cosmology and the more general and interesting framework with no $Z_2$ symmetry are under study.

\section*{Acknowledgments}
This research has been co-financed by the European Union (European 
Social Fund – ESF) and Greek national funds through the Operational 
Program "Education and Lifelong Learning" of the National Strategic 
Reference Framework (NSRF) - Research Funding Program: THALES. Investing 
in knowledge society through the European Social Fund.
We also acknowledge  the Research Special Account of the National and Kapodestrian University of Athens for partial financial supporting. \\


\end{document}